\documentclass[doublecol]{epl2}
\usepackage{graphicx}

\newcommand{\bk}{{\bf k}}

\newcommand{\bp}{{\bf p}}

\newcommand{\eps}{\epsilon}

\DeclareMathAlphabet{\mathpzc}{OT1}{pzc}{m}{it} 

\title{Comment on "Minimal conductivity in graphene: Interaction
corrections and ultraviolet anomaly" by Mishchenko E. G.}

\author{Igor F. Herbut\inst{1} \and Vladimir Juri\v ci\' c\inst{1}
\and Oskar Vafek\inst{2} \and Matthew J.  Case\inst{2}}

\institute{
  \inst{1} Department of Physics, Simon Fraser
University, Burnaby, British Columbia, V5A 1S6,Canada\\
\inst{2} National High Magnetic Field Laboratory and Department of
Physics, Florida State University, Tallahassee, Florida 32306, USA
} 
\abstract{}
\begin{document}

\maketitle

In a recent Letter\cite{Mishchenko08}, Mishchenko has presented a
calculation of the minimal conductivity of the clean graphene
taking into account the $1/r$ Coulomb interactions to first order
in perturbation theory, and questioned the results obtained
earlier by us\cite{Herbut08}. He performed this calculation in
three different ways and obtained three different results unless,
as he claimed, a proper short distance cutoff procedure
restricting large momentum transfers is implemented.

In this comment we wish to clarify and correct some of the
statements made in \cite{Mishchenko08}.

First, contrary to the claims in \cite{Mishchenko08} the
conductivity calculated using the Kinetic equation is identical to
the conductivity calculated from the Kubo formula. When
calculating transport properties, it is crucial to ensure that the
conservation laws are built into the approximation used for the
vertex. For the self-energy calculated at the Hartree-Fock level,
the four particle correlators must be calculated by summing the
RPA bubble and ladder diagrams (see Fig. 1 in
\cite{Mishchenko08}). Such non-perturbative treatment, while
strictly speaking uncontrolled, does reproduce the leading order
perturbative expansion and does not violate Ward identities
associated with the standard conservation laws. In general,
however, it does violate RG scaling laws. To calculate the
electrical conductivity, it is useful to define (in the reciprocal
space)
\begin{equation}
\Lambda_\mu({\bf k};i\Omega_n)\equiv\frac1\beta\sum_{i\omega_n}
\Pi_\mu(\bk,i\omega_n;\bk,i\omega_n+i\Omega_n),
\end{equation}
where $\Omega_n=2n\pi/\beta$ is a bosonic Matsubara frequency and
$\beta=1/T$, and the vertex matrix in the real space (and imaginary
time) is
\begin{eqnarray}
\Pi_{\mu}(x-y,y-x')=-\left\langle
T_{\tau}\psi(x)\psi^{\dagger}(x')J_{\mu}(y)\right\rangle.
\end{eqnarray}
Setting $e=v_F=\hbar=1$, the Dirac particle electrical current
operator is $J_{\mu}(y)=\psi^{\dagger}(y)\sigma_{\mu}\psi(y)$.
Within the previously mentioned approximation we have
\begin{eqnarray}
\label{VE} &&\Lambda_\mu({\bf
k};i\Omega_n)=-\frac1\beta\sum_{i\omega_n}
G_{\bk}(i\omega_n)\sigma_\mu
G_{\bk}(i\omega_n+i\Omega_n)\nonumber\\
&&-\int\frac{{\rm d}^2{\bf p}}{(2\pi)^2}V_{\bf k-p}\frac1\beta
\sum_{i\omega_n}G_{\bk}(i\omega_n)\Lambda_\mu({\bf p};i\Omega_n)
G_{\bk}(i\omega_n+i\Omega_n).\nonumber\\
\end{eqnarray}
The above corresponds to $4$ coupled integral equations. A rather
straightforward calculation shows that the vertex matrix function
$\Lambda_{\mu}(\bk,i\Omega_n)$ can be written in the following
form
\begin{eqnarray}
\Lambda_\mu({\bf k};i\Omega)=
\left(\delta_{\mu\nu}-\hat{k}_\mu\hat{k}_\nu\right)\sigma_\nu
f_{\bk}(i\Omega)+\epsilon_{\mu\nu}\hat{k}_\nu\sigma_3g_{\bk}(i\Omega).
\end{eqnarray}
Here, $\mu$ runs from $1$ to $2$ and the sum over repeated indices
is understood; $\eps_{\mu\nu}$ is the completely antisymmetric
rank 2 tensor and $\hat{k}$ is a unit length wavevector. This
decomposition allows us to reduce the set of $4$ coupled integral
equations to a set of $2$ coupled integral equations for $f$ and
$g$ with a non-separable vertex. At $T=0$, this set of equations
reads
\begin{eqnarray}\label{inteq}
\label{TE}&& \left[
\begin{array}{cc}
2\epsilon_{\bf k}&-\Omega\\
\Omega&2\epsilon_{\bf k}
\end{array}
\right]\left[
\begin{array}{c}
f_{\bk}(i\Omega)\\
g_{\bf k}(i\Omega)
\end{array}
\right]=\left[
\begin{array}{c}
1\\0
\end{array}
\right]+\nonumber\\
&+&\int\frac{{\rm d}^2{\bf p}}{(2\pi)^2} \hat{\bf k}\cdot\hat{\bf
p} V_{\bf k-p}\left[
\begin{array}{c}
\hat{\bf k}\cdot\hat{\bf p} f_{\bp}(i\Omega)\\ g_{\bp}(i\Omega)
\end{array}
\right]
\end{eqnarray}
where the Hartree-Fock energy is given by
\begin{equation}
\epsilon_{\bf k}=k+\frac12\int\frac{{\rm d}^2{\bf p}}{(2\pi)^2}
\hat{\bf k}\cdot\hat{\bf p}V_{\bf k-p},
\end{equation}
and (restoring the physical units) the Coulomb potential is
\begin{eqnarray}
V_{\bk}=\frac{2\pi}{\eps}\frac{e^2}{\hbar v_F}\frac{1}{|\bk|}.
\end{eqnarray}
Eqs. (\ref{TE}) constitutes the leading approximation to the
quantum transport theory for massless Dirac fermions in the
collisionless limit. The electrical conductivity\footnote{ The
conductivity can be calculated without analytical continuation by
subtracting the zero frequency component of $f$ before taking the
limit $\Omega\rightarrow0$:
$$
\sigma=-2\pi\lim_{\Omega\rightarrow0}\frac1{\Omega}\int\frac{{\rm
d}^2{\rm k}}{(2\pi)^2}\left\{f({\bf k};i\Omega)-f({\bf
k};0)\right\}.
$$} for $N$ two components Dirac flavors, in units of $Ne^2/h$, can be calculated using
\begin{eqnarray}
\sigma=-2\pi\lim_{\Omega\rightarrow0}\frac1{\Omega}\int\frac{{\rm
d}^2{\rm k}}{(2\pi)^2}\Im m f({\bf k};\Omega+i0^+).
\end{eqnarray}
The Figure shows the numerical result found perturbatively for
small $\alpha=e^2/(\epsilon\hbar v_F)$ which we compare with the
analytic result found to first order in $\alpha$.
\begin{figure}
\onefigure[width=0.45\textwidth]{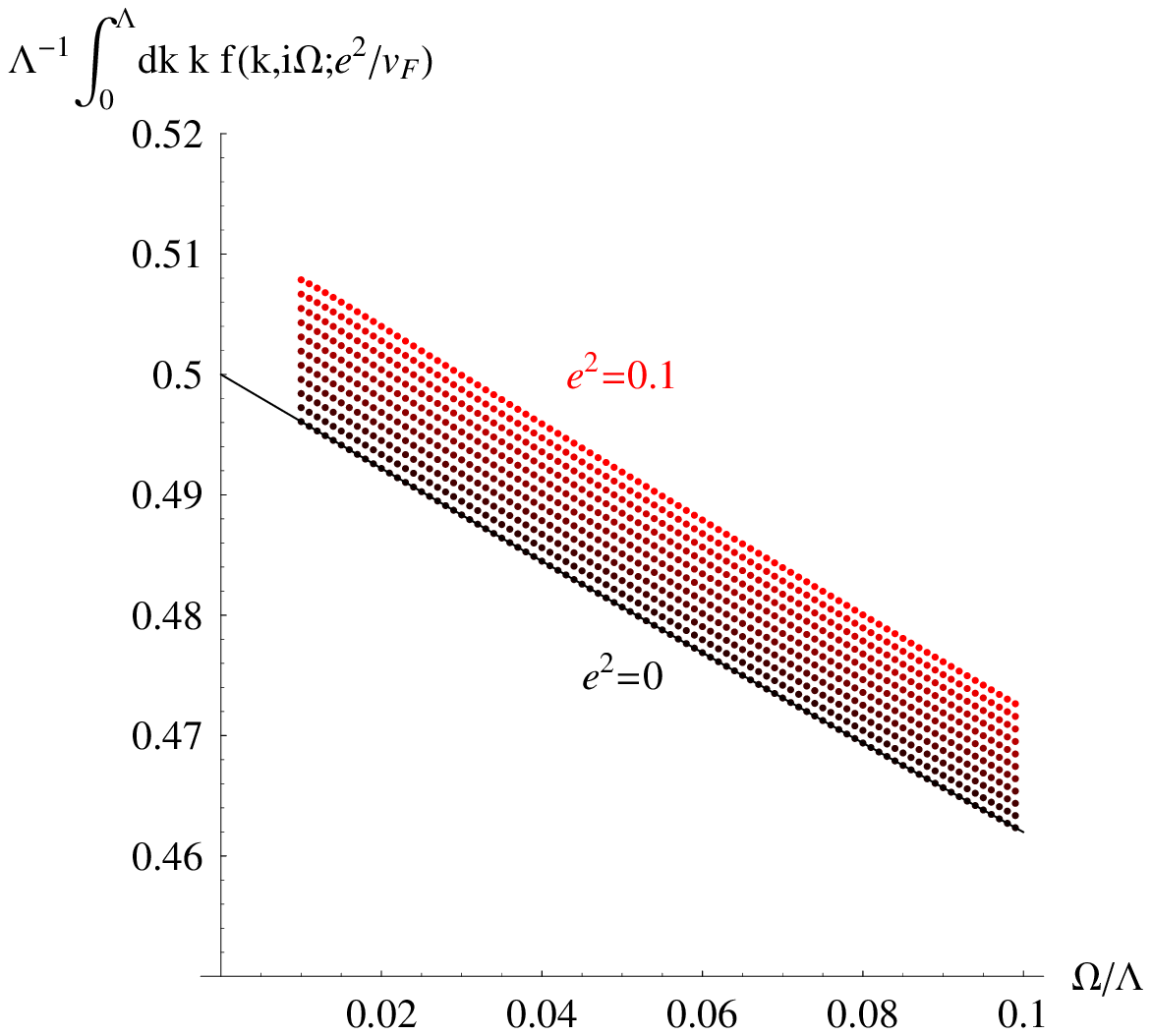}
\onefigure[width=0.45\textwidth]{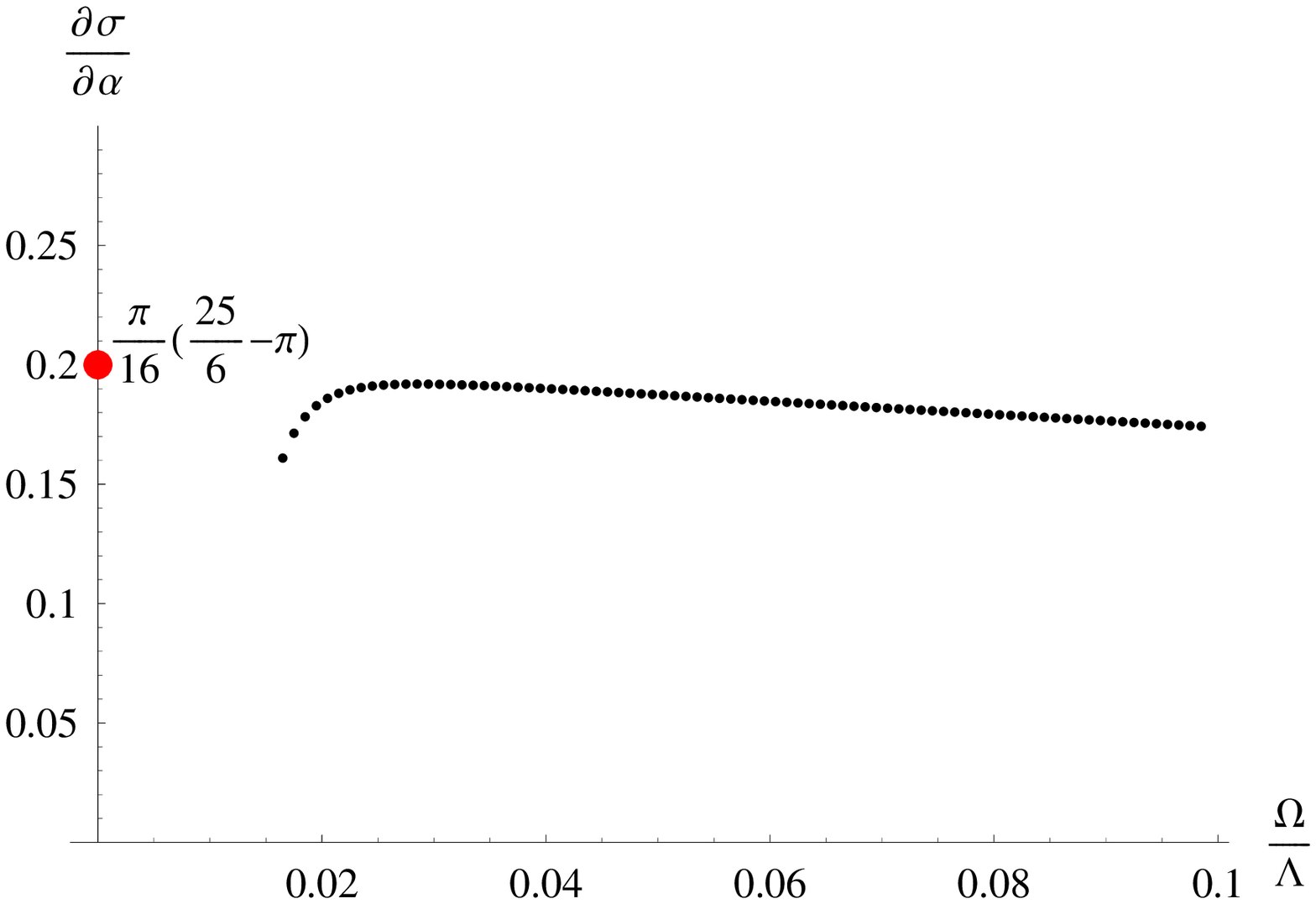}
\caption{(Upper panel) This is the result of the numerical
solution of the integral equation (\ref{inteq}), integrated over
$dk k$ for different values of $e^2$, ($v_F=1,\eps=1$), and
(Euclidean) $\Omega$. The analytical result for the $e^2=0$,
($\frac{1}{2}-\frac{\Omega}{4}\tan^{-1}\frac{2}{\Omega}$) is also
plotted with a solid line. The agreement at $e^2=0$ is to the
$6th$ significant digit. The $\Omega$-slope of these lines is
found at different values of $\Omega$ and plotted vs. $e^2$ (not
shown). (Lower panel) The $\alpha$-slope is then plotted for
different values of $\Omega$. The red dot is the value found
analytically. The deviation at small frequency is likely a result
of numerical errors.} \label{pivsomega}
\end{figure}

Second, and more importantly, while the conductivity calculated
using Kubo formula would superficially indeed seem to depend on
the cutoff procedure, further analysis shows, of course, that it
does not. This point can be amply illustrated by using {\it two}
different cutoffs: one hard cutoff $\Lambda$ which restricts the
fermion modes to the vicinity of the Dirac point, and second $M$
which restricts the momentum transfer, i.e.,
\begin{eqnarray}
V_{\bk}\rightarrow \frac{2\pi}{\eps}\frac{e^2}{\hbar
v_F}\left(\frac{1}{|\bk|}-\frac{1}{\sqrt{\bk^2+M^2}}\right).
\end{eqnarray}
In the high energy parlance the second way of cutting off the
theory corresponds to the Pauli-Villars regularization. The result
obtained using this procedure is the dc conductivity
\begin{equation}
\sigma=N\frac{e^2}{h}\left[\frac{\pi}{8}+
\alpha\frac{\pi}{16}\left(\frac{25}{6}-\pi-\frac{1}{\sqrt{1+\frac{M^2}{\Lambda^2}}}\right)\right]
\end{equation}
where $\alpha=e^2/\hbar v_F$ and $N=4$ is the number of fermion
flavors in graphene. Naively, this would seem to lead to a one
parameter family of results that interpolate between our original
result (recovered when $M/\Lambda\rightarrow \infty$) and the one
advocated by Mishchenko (when $M/\Lambda\rightarrow 0$). The root
of this apparent ambiguity may be  easily understood by
considering the limit $M \gg \Lambda$, for example. One finds,
$$
\sigma \approx
\frac{e^2}{h}\left[\frac{\pi}{8}+\alpha\frac{\pi}{16}\left(\frac{25}{6}-\pi-\frac{\Lambda}{M}\right)\right].
$$
The last term proportional to $\alpha \Lambda/M$ thus corresponds
to the perturbatively {\it irrelevant} short range part of the
interaction. Indeed, if one would compute the correction to the
Gaussian conductivity in the same system of Dirac fermions
interacting via short-range interactions only, the result would be
$$
\sigma'(\Omega) = \frac{1}{\Omega} [ a \lambda \Lambda^2 + b
\lambda \Lambda  \Omega
 + c \lambda \Omega^2 + O(\Omega^3, \lambda^2)] \frac{e^2}{h}
$$
where $a$, $b$, and $c$ are numerical constants. $\lambda$ is the
dimensionfull short range interaction coupling constant with the
dimension one, that can be identified with $\sim \alpha/M$ in the
preceding discussion. The first {\it two} terms explicitly contain
the cutoff, and thus clearly represent artifacts of the
calculation that violates gauge invariance. {\it Both} of these
terms must be dropped, so that the final result becomes
$$
\sigma' (\Omega) = [c \lambda \Omega +O(\Omega^2, \lambda^2)]
\frac{e^2}{h}.
$$
This then agrees with the irrelevance of the short range
interaction by power counting. The same result is obtained using
the dimensional regularization of Veltman and t'Hooft, which was
designed precisely to automatically discard the spurious terms of
the type of the above.

The very last term in Eq. (9) therefore is the consequence of the
incorrect inclusion of the irrelevant short range interaction into
the leading correction to the Gaussian result. Such a leading
correction can come only from the single marginally irrelevant
coupling in the theory, which is the Coulomb interaction. Short
range interaction provides only the further, sub-leading
corrections, proportional to the frequency. Kubo formula this way
leads to the unique result obtained previously by
us\cite{Herbut08}.

The last point raised by Ref.\cite{Mishchenko08} is that the
conductivity calculated using density-density correlation function
followed by the continuity equation is divergence free and
coincides with the Pauli-Villars result. While there is indeed no
logarithmic divergence in vertex correction diagram, the self
energy is logarithmically divergent, as it may be readily seen
from Eq.\ (12) in Ref.\ \cite{Mishchenko08}, and must be cut off.
We admit that the agreement between these two procedures is
puzzling. Furthermore, the Coulomb correction to the dc
conductivity calculated using the density-density correlator is
not unique, i.e., it depends on the procedure implemented to
regularize integrals. Namely, the self-energy diagram when
calculated within the dimensional regularization scheme, together
with the vertex diagram, yields the correction to the conductivity
$\sigma_{\rm pol}^{\rm dim-reg}=\sigma_0\alpha(11-3\pi)/6$, {\it
different} than one obtained in \cite{Mishchenko08} using sharp
momentum cutoff. Here $\sigma_0=\pi/2$ is the noninteracting
(Gaussian) conductivity in units $e^2/h$. The origin of this
non-uniqueness is unclear at the moment, but we suspect that it
may be the non-gauge invariant contribution to the conductivity
which is not properly treated within the density polarization
approach. On the other hand, as we have already discussed, the
Kubo formula yields a unique, regularization independent Coulomb
correction to the dc conductivity found by us in Ref.\
\cite{Herbut08}.

\end{document}